\documentclass{Interspeech}

\interspeechcameraready

\usepackage{xcolor}
\usepackage{makecell}

\title{Towards General Discrete Speech Codec for Complex Acoustic Environments: A Study of Reconstruction and Downstream Task Consistency}

\author[affiliation={1}]{Haoran}{Wang}
\author[affiliation={1}]{Guanyu}{Chen}
\author[affiliation={1}]{Bohan}{Li}
\author[affiliation={1}]{Hankun}{Wang}
\author[affiliation={1}]{Yiwei}{Guo}
\author[affiliation={1}]{Zhihan}{Li}
\author[affiliation={1}]{Xie}{Chen}
\author[affiliation={1}]{Kai}{Yu}

\affiliation{MoE Key Lab of Artificial Intelligence, Jiangsu Key Lab of Language Computing, X-LANCE Lab, School of Computer Science}{Shanghai Jiao Tong University}{China}
\email{wanghaor@sjtu.edu.cn, kai.yu@sjtu.edu.cn}
\keywords{neural codec, speech processing, speech recognition, speech enhancement}

\usepackage{comment}
\usepackage{algorithm}
\usepackage{algorithmic}
\usepackage{multirow}
\usepackage{subfig}
\definecolor{DeepGreen}{RGB}{0,135,62}
\definecolor{PumpkinOrange}{RGB}{153,102,204}    
\definecolor{Amethyst}{RGB}{255,117,24}         
\begin{document}

\maketitle

\begin{abstract}
Neural speech codecs excel in reconstructing clean speech signals; however, their efficacy in complex acoustic environments and downstream signal processing tasks remains underexplored. In this study, we introduce a novel benchmark named \textbf{E}nvironment-\textbf{R}esilient \textbf{S}peech Codec \textbf{B}enchmark (ERSB) to systematically evaluate whether neural speech codecs are \emph{environment-resilient}. Specifically, we assess two key capabilities: (1) robust reconstruction, which measures the preservation of both speech and non-speech acoustic details, and (2) downstream task consistency, which ensures minimal deviation in downstream signal processing tasks when using reconstructed speech instead of the original. Our comprehensive experiments reveal that complex acoustic environments significantly degrade signal reconstruction and downstream task consistency. This work highlights the limitations of current speech codecs and raises a future direction that improves them for greater environmental resilience.
\end{abstract}
\section{Introduction}
{\let\thefootnote\relax\footnotetext{\vspace{-0.1in}Kai Yu is the corresponding author.}}
Neural speech codecs play a crucial role in addressing the increasing demand and diverse tasks related to speech processing. 
These codecs typically consist of an encoder, a quantization module, and a decoder. 
The speech signal is encoded into a sequence of discrete codes through the encoder and quantizer, which can then be used for efficient data transmission, storage, or directly serve as input or targets for various downstream tasks~\cite{valle, wang2024viola, yao2025gense}. 
To meet these demands, a fundamental requirement for neural speech codecs is to preserve as much information from the original speech signal as possible in the codes, enabling high-quality reconstruction through the decoder.

However, although research on neural speech codecs has rapidly expanded in recent years~\cite{encodec, kumar2024high, zhang2024speechtokenizer, liu2024semanticodec, ye2024codec} (see also an overview~\cite{guo2025review}), many of them are trained and tested only on clean speech datasets. 
In contrast, real-world acoustic environments are highly complex, with numerous long-tail scenarios (e.g., stationary or non-stationary background noise or music of varying intensity). 
In addition, many real-world applications such as online conferencing systems and translation services heavily rely on \emph{signal processing systems} whose input is continuous audio signals. 
In these scenarios, a speech signal in complex acoustic environments is first compressed to codes by a codec, then transmitted, decoded to waveforms, and fed to backend signal processing systems like speech enhancement and recognition.
To better retain information and facilitate downstream signal processing tasks, a general codec should be \emph{environment-resilient} that possesses two key capabilities, even in \emph{complex acoustic environments}:
\begin{enumerate}
    \item \textbf{Robust Reconstruction Quality}: the ability to preserve not only speech intelligibility and quality but also various non-speech acoustic details in the reconstructed audio.
    \item \textbf{Downstream Task Consistency}: the ability to maintain reconstruction equivalence, i.e. the reconstructed speech, when used as input for downstream signal processing tasks, causes minimal deviation compared to directly using the original speech.
\end{enumerate}
Therefore, establishing a comprehensive evaluation methodology for codec performance in such environments is essential. This requires assessing these two key environment-resilient capabilities in diverse and complex acoustic conditions. Note that semantic tokenizers such as HuBERT~\cite{hsu2021hubert} and $\mathcal S^3$ Tokenizer~\cite{du2024cosyvoice} are beyond the scope of this paper since they are neither designed for compression and reconstruction nor can they serve as input to signal processing systems.

Although several codec benchmark studies have been proposed, their evaluations are generally conducted on clean speech data. 
Codec-SUPERB~\cite{wu2024codec} is a public benchmarking framework that provides a comprehensive evaluation of neural speech codecs, integrating multiple datasets and diverse metrics. However, its datasets and tasks are based on clean speech, pure music, and pure audio sound data.
DASB~\cite{mousavi2024dasb} categorizes evaluation tasks into discriminative and generative types. Discriminative tasks include speech recognition, speaker identification, etc., while generative tasks encompass text-to-speech, speech enhancement, etc.  
Although DASB incorporates noisy speech data in the speech enhancement task, it directly feeds discrete tokens into downstream models, and only evaluates how well the enhanced speech approximates the target clean speech. 
This does not indicate whether the codec can \emph{reconstruct} noisy speech with robust quality, let alone assess the impact of the reconstruction process on downstream tasks.  

In this paper, we propose a novel benchmark named \textbf{E}nvironment-\textbf{R}esilient \textbf{S}peech Codec \textbf{B}enchmark (ERSB) to evaluate whether a neural speech codec is environment-resilient. We design a pipeline to assess various codecs in terms of the two key capabilities mentioned earlier under complex acoustic environments. For reconstruction quality, we conduct evaluations using multiple datasets, incorporating both simulated and real-world data to test various reconstruction metrics. For downstream task consistency, we assess the impact of codec-reconstructed audio on downstream model outputs in two signal processing backends: speech enhancement (SE), and automatic speech recognition (ASR) with SE. 
The main contributions of our paper can be summarized as:
\begin{enumerate}
    \item We raise an important new issue: the environmental resilience of neural speech codecs. We then practically decompose it into two key capabilities: robust reconstruction quality and consistency on downstream signal processing tasks.
    \item We propose a new benchmark with various datasets and a novel pipeline to comprehensively evaluate the environmental resilience of neural speech codecs. 
    Using this benchmark, several mainstream codecs are evaluated.
    \item The experiments reveal that most of the tested codecs perform poorly in terms of reconstruction in complex acoustic environments.
    While a few codecs have comparatively better reconstruction quality, such as DAC~\cite{kumar2024high}, they still lack downstream task consistency. 
    Our findings suggest that further research is needed to develop environment-resilient codecs.
\end{enumerate}

\section{Environment-Resilient Speech Codec Benchmark}
In this section, we first present our data simulation process in Section \ref{s2.1}, which plays a crucial role in constructing the benchmark dataset with noisy data from the real world, ensuring a diverse range of acoustic environments. Subsequently, we introduce the framework for evaluating environmental resilience in Section \ref{s2.2}, which is divided into an assessment of reconstruction quality and performance consistency evaluations of two downstream signal processing systems.

\begin{figure}[t]
  \centering
  \includegraphics[width=\linewidth]{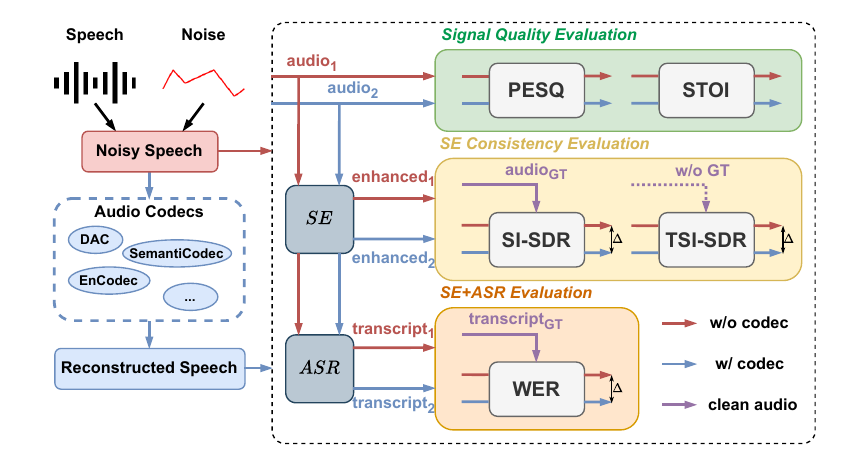}
  \caption{Illustration of ERSB benchmark framework.}
  \label{fig:figure1}
\end{figure}
\subsection{Complex Acoustic Environment Data Simulation}
\label{s2.1}
Following DNS challenge 5~\cite{dns5}, the data simulation process incorporates clean speech, noise, and room impulse responses. This process can be formally expressed as:
\begin{equation}
    \mathcal{M}(t) = \mathcal{I}(t) * \mathcal{S}(t) + \mathcal{N}(t)
\end{equation}
where $t$ denotes the time index of the mixture signal $ \mathcal{M} $, the clean speech signal $\mathcal{S}$, the room impulse response $\mathcal{I}$, and the noise signal $\mathcal{N}$. Root mean square (RMS) normalization is applied to each signal to ensure consistency in amplitude levels.

To enhance the diversity of the simulated data, we regulate both the signal-to-noise ratio (SNR) and the loudness of the generated signal. This relationship is mathematically represented as:
\begin{equation}
    \mathcal{M}(t) = 10^{l/20}(\mathcal{I}(t) * \mathcal{S}(t) + 10^{-\mu/20}\mathcal{N}(t))
\end{equation}
where $\mu$ denotes the signal-to-noise ratio (SNR), and $l$ represents the loudness.


\subsection{Codec Environmental Resilience Evaluation}
\label{s2.2}
Speech codec acoustic environmental resilience refers to the ability of a speech codec to maintain consistent performance in signal reconstruction and backend processing under varying acoustic conditions. To assess this resilience, we conduct evaluations using both simulated and real-world data, focusing primarily on signal reconstruction quality and the performance of two commonly considered backend processing scenarios.

\subsubsection{Signal Reconstruction Quality}
For signal reconstruction evaluation, we assess the waveform-level fidelity between the original speech signals from the dataset and their corresponding reconstructions generated by various speech codecs. To quantify reconstruction quality, we employ two widely recognized standardized metrics: Perceptual Evaluation of Speech Quality (PESQ) and Short-Time Objective Intelligibility (STOI). In this context, the original speech signals serve as reference inputs, while the reconstructed signals—obtained by encoding and subsequently decoding the original signals through different codecs—are the subjects of measurement. Under a specific acoustic condition, higher metrics refer to better codec performance of reconstruction.
\begin{figure*}[h]
  \centering
  \includegraphics[width=\linewidth]{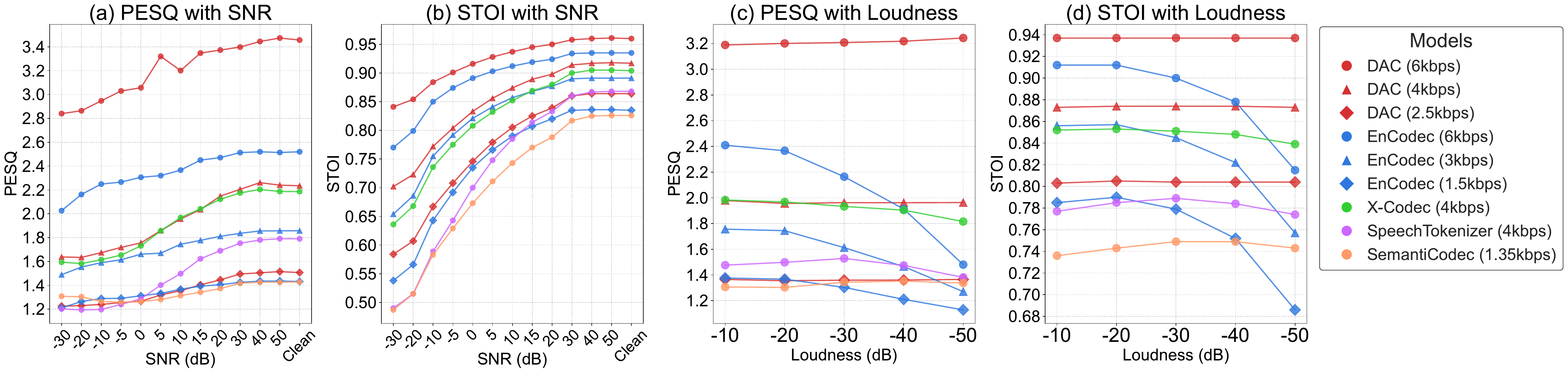}
  \caption{PESQ/STOI values with respect to SNR and loudness.}
  \label{fig:figure2}
\end{figure*}
\subsubsection{Consistency of Signal Quality after Enhancement}
In practical applications, most real-world speech signals are inherently noisy, necessitating the incorporation of a SE module within speech signal processing systems. To assess the consistency of SE performance, we apply speech enhancement to both the audio before and after codec reconstruction, yielding $\textit{enhanced}_{1}$ and $\textit{enhanced}_{2}$, respectively The clean audio, which is used during the synthesis process, serves as the ground truth ($\textit{audio}_{\textit{GT}}$). We compare \(\textit{enhanced}_{1}\) and \(\textit{enhanced}_{2}\) with $\textit{audio}_{GT}$ to obtain Scale-Invariant Signal-to-Distortion Ratio (SI-SDR), \(\textit{SI-SDR}_{1}\) and \(\textit{SI-SDR}_{2}\), respectively. 
The signal loss due to the reconstruction process under the speech enhancement is determined by the difference: $\Delta \textit{SI-SDR} = \textit{SI-SDR}_{2}-\textit{SI-SDR}_{1}$.

For real-world datasets, clean audio is not available as a reference for the SI-SDR calculation. TorchAudio-Squim~\cite{TorchAudio-Squim} provides a reliable method to predict an approximate SI-SDR without a reference, which is denoted as TSI-SDR and utilized for evaluating noise levels in real-world data in this work. The $\Delta$\textit{TSI-SDR} represents the difference in TSI-SDR values directly calculated from the enhanced audio before and after reconstruction:$\Delta \textit{TSI-SDR} = \textit{TSI-SDR}_{2}-\textit{TSI-SDR}_{1}$.

\subsubsection{Consistency of Speech Recognition after Enhancement} 

Signal processing systems are predominantly realized through cascaded architectures. Consequently, we have constructed a straightforward signal processing system comprising a SE module followed by an ASR module. In our evaluation, the clean speech transcription serves as the ground truth ($\textit{transcript}_{\textit{GT}}$), while we obtain $\textit{transcript}_1$ from the ASR transcription of the enhanced original audio and $\textit{transcript}_2$ from the ASR transcription of the enhanced codec-reconstructed audio.

To quantify performance differences, we compute the WER for each case: $\textit{WER}_1$ for $\textit{transcript}_1$ against $\textit{transcript}_{\textit{GT}}$, and $\textit{WER}_2$ for $\textit{transcript}_2$ against $\textit{transcript}_{\textit{GT}}$. The performance variation of the pipeline due to codec reconstruction process is then determined by the difference: $\Delta{\textit{WER}} = \textit{WER}_2 - \textit{WER}_1$.
A higher $\Delta{\textit{WER}}$ signifies greater degradation in ASR performance introduced by codec reconstruction in conjunction with speech enhancement, highlighting the codec’s influence on the robustness of this signal processing system.  
\section{Experiment}


\subsection{Setup}
To build the dataset, we make modifications based on the audio simulation script in DNS Challenge 5~\cite{dns5}. The clean audio used for simulation comes from the read speech in DNS Challenge 5~\cite{dns5}. The pure noise data comes from the lounge noise recorded in CHiME1~\cite{chime1}, which, on top of white noise, contains various ambient sounds such as television noise, the laughter of children, and indistinct conversations. The room impulse response (RIR) comes from the RIR datasets of DNS Challenge 5~\cite{dns5}. Two experimental variables are controlled:


\begin{itemize}
    
    \item \textbf{SNR variation}: We fix loudness at -20 dB and the RIR, then adjust the SNR from -30 dB to 50 dB. The lower the SNR, the higher the proportion of noise, making the audio sound noisier.
    
    \item \textbf{Loudness variation}: We fix the SNR at -10 dB and the RIR, then adjust the loudness from -50 dB to -10 dB. The smaller the loudness value, the quieter the audio sounds.
\end{itemize}

Additionally, for the real-world datasets, we select the real record part from the DNS Challenge 1~\cite{dns1} blind test set and the real part of CHiME4~\cite{vincent20164th}. For real-world audio dataset CHiME4, given the inherent noise in the audio, we start by assessing the noise level using the TorchAudio-Squim~\cite{TorchAudio-Squim} to measure an approximate SI-SDR without a reference, which is denoted as TSI-SDR.

In terms of codecs, we select the following codecs as test objects: DAC~\cite{kumar2024high} (6 kbps, 4 kbps, 2.5 kbps) and EnCodec~\cite{encodec} (6 kbps, 3 kbps, 1.5 kbps) as two famous general-purpose codecs;  SemantiCodec~\cite{liu2024semanticodec} (1.35 kbps), SpeechTokenizer~\cite{zhang2024speechtokenizer} (4 kbps), and X-Codec~\cite{ye2024codec} (4 kbps) as three typical codecs with semantic distillation~\cite{guo2025review}. 
We use the official checkpoints for all these codecs.
The composition of the training sets for each codec model can be found in Table \ref{table:table1}. For DAC, 4 kbps and 2.5 kbps models are obtained by taking the first 8 layers and 5 layers of the 12-layer residual vector quantization of DAC respectively.
\begin{table}[h]
    \centering
    \footnotesize
    \caption{Composition of training data. ``-'' means not reported in the paper or the open-source project whether the model has data of this category involved in the training.}
    \label{table:table1}
    \begin{tabular}{lccc}
        \toprule
        \textbf{Codec} & \textbf{Clean Speech} & \textbf{Music} &\textbf{ General Sound} \\
        \midrule
        DAC & $\checkmark$ & $\checkmark$ & $\checkmark$ \\
        EnCodec & $\checkmark$ & $\checkmark$ & $\checkmark$ \\
        SemantiCodec & $\checkmark$ & $\checkmark$ & $\checkmark$ \\
        SpeechTokenizer & $\checkmark$ & $\times$ & $\times$ \\
        X-Codec & - & - & $\checkmark$ \\
        \bottomrule
    \end{tabular}
\end{table}

For the ASR task, we employ the Whisper~\cite{whisper} Large-v3 model. 
For speech enhancement, we evaluated multiple models and selected the best-performing one: SepFormer~\cite{subakan2021attention}, trained on the WHAM! dataset~\cite{wham}. For the evaluation of signal metrics, we select the VERSA toolkit~\cite{shi2024versa} to measure the PESQ, STOI, and SI-SDR metrics.

\begin{table}[h]
  \caption{Codec reconstruction evaluations on real record part from the DNS Challenge 1 blind test set (corresponding to DNS 1 in the table) and the real part of CHiME4.}
  \label{table:table2}
  \centering
  \footnotesize
  \setlength{\tabcolsep}{4.2pt}
  \renewcommand{\arraystretch}{1.15} 
  \begin{tabular}{ccccccc}
  \toprule
  \multirow{2}{*}{\textbf{Codec}} & \multirow{2}{*}{\makecell{\textbf{Bitrate}\\\textbf{(kbps)}}} & \multicolumn{2}{c}{\textbf{PESQ}$\uparrow$} & \multicolumn{2}{c}{\textbf{STOI}$\uparrow$} \\ 
  \cline{3-4} \cline{5-6}
   &  & DNS1 & CHiME4 & DNS1 & CHiME4 \\
\hline

  SemantiCodec & 1.35 & 1.66 & 1.37 & 0.77 & 0.59 \\ 
\hline
  SpeechTokenizer & 4.00 & 1.94 & 1.58 & 0.82 & 0.67 \\ 
\hline
  X-Codec & 4.00 & 2.43 & 1.97 & 0.87 & 0.77 \\ 
\hline

  \multirow{3}{*}{EnCodec} & 1.50 & 1.42 & 1.46 & 0.77 & 0.69 \\ 

   & 3.00 & 1.85 & 1.94 & 0.84 & 0.80 \\ 

    & 6.00 & 2.44 & 2.80 & 0.89 & 0.89\\ 
   \hline
\multirow{3}{*}{DAC}  & 2.50 & 1.63 & 1.49 & 0.82 & 0.71\\ 

   & 4.00 & 2.40 & 2.40 & 0.89 & 0.83 \\ 

    & 6.00 & 3.63 & 3.48 & 0.94 & 0.90\\ 
\bottomrule
\end{tabular}
\end{table}

\begin{table*}[ht]
\centering
\footnotesize
\setlength{\tabcolsep}{6pt}
\renewcommand{\arraystretch}{1.15} 
\caption{$\Delta$TSI-SDR (dB) of SE task and $\Delta$WER (\%) of SE+ASR task on the TSI-SDR-divided CHiME4. For TSI-SDR-divided CHiME4, TorchAudio-Squim predicts approximate SI-SDR (TSI-SDR) per audio for dataset division by TSI-SDR ranges. \textbf{ Case A, B, C, and D represent the the dataset division of SI-SDR \textless{} -5 dB, -5 $\sim$ 0 dB, 0 $\sim$ 5 dB, and \textgreater{} 5 dB, respectively}. ``w/o codec recon.'' means no codec reconstruction; corresponding row data shows TSI-SDR and WER (baseline), not $\Delta$TSI-SDR and $\Delta$WER.}
\label{table:table3}
\begin{tabular}{c|c|cccc|cccc}
\toprule
\multirow{2}{*}{\textbf{Codec}} & \multirow{2}{*}{\textbf{\makecell{Bitrate\\(kbps)}}} & \multicolumn{4}{c|}{\textbf{$\Delta$ TSI-SDR (dB)$\uparrow$}} & \multicolumn{4}{c}{\textbf{$\Delta$ WER ($\%$)$\downarrow$}} \\ \cline{3-10}
    & & \multirow{1}{*}{\textbf{Case A}} & \multirow{1}{*}{\textbf{Case B}} & \multirow{1}{*}{\textbf{Case C}} & \multirow{1}{*}{\textbf{Case D}} & \multirow{1}{*}{\textbf{Case A}} & \multirow{1}{*}{\textbf{Case B}} & \multirow{1}{*}{\textbf{Case C}} & \multirow{1}{*}{\textbf{Case D}} \\ \hline
w/o codec recon. & - & 16.10 & 17.69 & 19.70 & 22.68 & 13.94 & 7.97 & 5.69 & 4.51\\ \cline{1-10}

SemantiCodec & 1.35 & 
-16.44 &  
-13.65 &  
-13.29 &  
-11.31 & 
+89.00 &  
+81.59 &  
+62.93 &  
+24.45\\ 
\cline{1-10}

SpeechTokenizer & 4.00 & 
-9.46 &  
-5.47 &  
-3.66 &  
-2.14 & 
+74.14 &  
+53.70 &  
+26.84 &  
+7.68\\  
\cline{1-10}

X-Codec & 4.00 & 
-5.09 &  
-3.14 &  
-2.73 &  
-1.89 & 
+47.52 &  
+32.42 &  
+16.28 &  
+5.21\\ \cline{1-10}  

\multirow{3}{*}{EnCodec} & 1.50 & 
-21.83 &  
-15.20 & 
-13.40 &  
-14.63 & 
+99.66 &  
+105.11 & 
+95.95 &  
+70.91\\ 

& 3.00 & 
-15.72 &  
-10.20 &  
-8.79 &  
-9.34 & 
+87.51 &  
+71.01 &  
+45.34 &  
+17.30 \\ 

& 6.00 & 
-8.78 &  
-5.43 &  
-4.54 &  
-4.13 & 
+55.54 &  
+31.42 &  
+13.94 &  
+4.12\\ \cline{1-10}  
\multirow{3}{*}{DAC} 
& 2.50 & 
-14.19 &  
-10.35 &  
-9.30 &  
-9.95 & 
+96.88 &  
+88.09 &  
+64.65 &  
+28.55\\ 

& 4.00 & 
-5.85 &  
-4.04 &  
-3.52 &  
-2.88 & 
+44.80 &  
+27.25 &  
+12.45 &  
+3.53\\  

& 6.00 & 
-3.42 &  
-2.18 &  
-1.72 &   
-0.97 & 
+27.65 &  
+14.48 &  
+6.30 &   
+1.86\\  

\bottomrule
\end{tabular}
\end{table*}

\subsection{Reconstruction Quality}
In the reconstruction quality test, we examine how the reconstruction quality of the mentioned codecs change when varying SNR and loudness. We visualize the results in Figure \ref{fig:figure2} to show PESQ/STOI values with SNR and loudness. 


As shown in Figure \ref{fig:figure2} (a) and (b), when the SNR is extremely low, most codecs, except 6kbps DAC and EnCodec, exhibit poor reconstruction performance with PESQ values below $2$, suggesting clear audible differences. And low-bitrate models, such as SemantiCodec(1.35kbps) and EnCodec(1.5kbps), perform poorly in reconstruction performance. However, since SpeechTokenizer has not been trained on noisy data, its performance improves significantly as the SNR increases. Moreover, Figure \ref{fig:figure2} (c) and (d) reveal that among the tested models, for all EnCodec models with various bitrates, as the loudness decreases, the reconstruction quality drops significantly.

In addition, we also conduct tests on the real record part from the DNS Challenge 1 blind test set and the real part of CHiME4. 
From Table \ref{table:table2}, we can observe the performance of each codec on real-world data. On real-world data, the performance of the codec is similar to that on simulated data.

\subsection{Performance of Signal Processing Backends}

\begin{figure}[t]
  \centering
  \includegraphics[width=\linewidth]{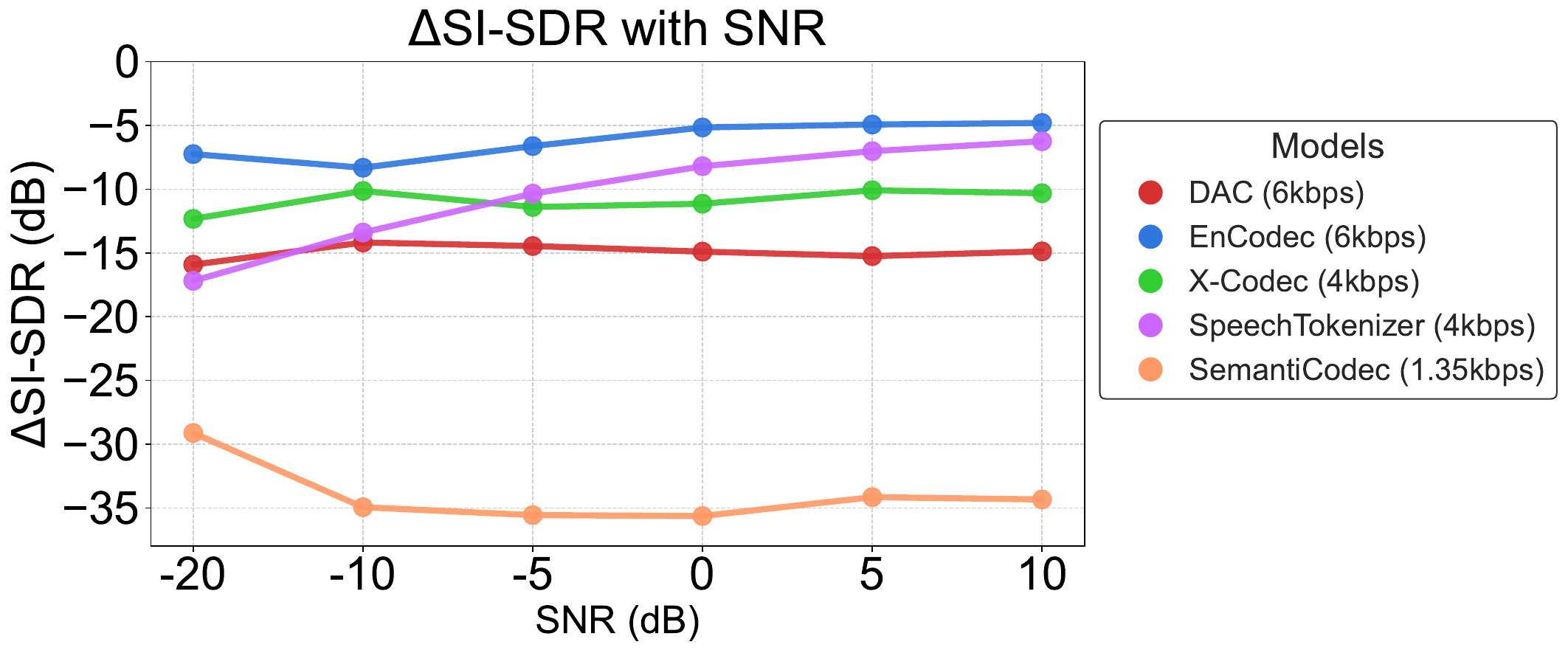}
  \caption{Consistency of the SE backend on the simulated dataset, measured by $\Delta$SI-SDR.}
  \label{fig:figure3}

  \vspace{0.1cm} 
  
  \includegraphics[width=\linewidth]{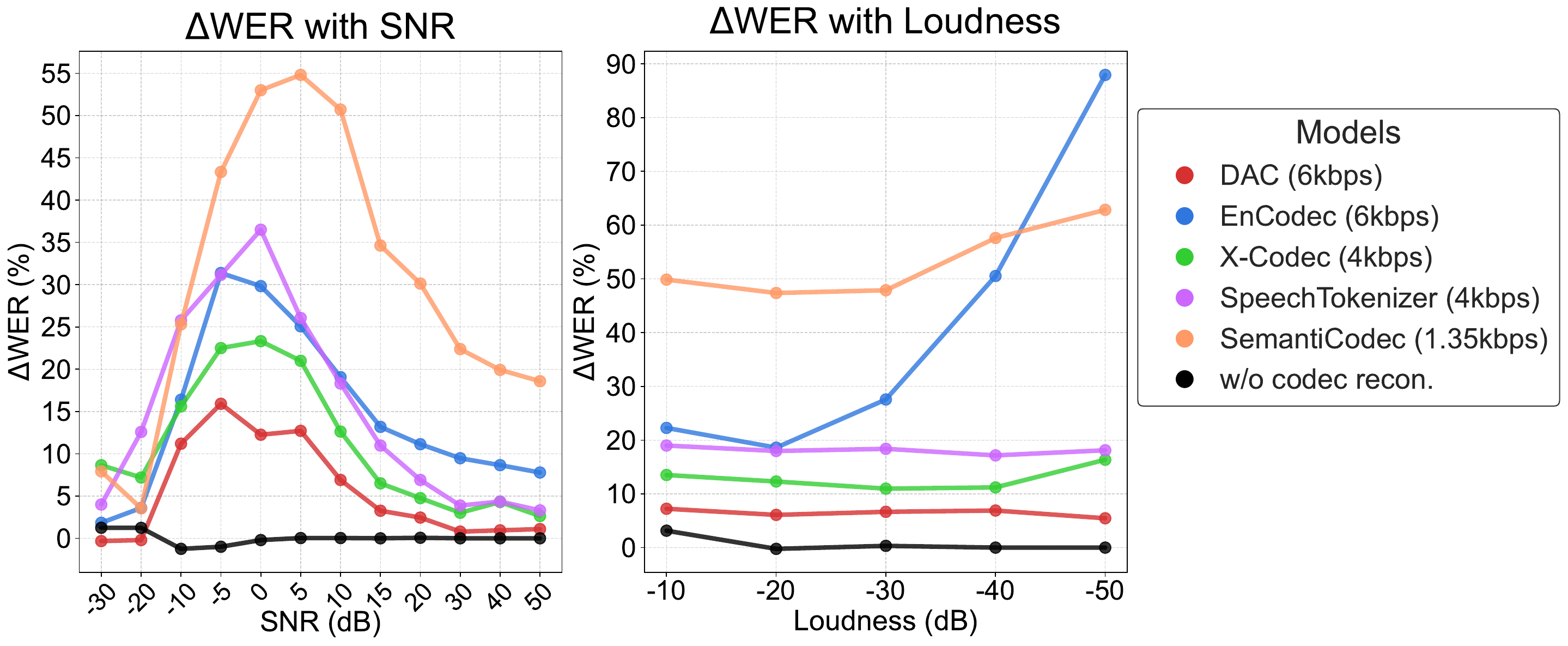}
  \caption{Consistency of the SE+ASR backend on the simulated dataset, measured by $\Delta$WER.}
  \label{fig:figure4}
  \vspace{-0.4cm}
\end{figure}




\subsubsection{Consistency of SE Backend}
First, we measure the variation of $\Delta$SI-SDR with SNR for different codecs after reconstruction and enhancement. 
From Figure \ref{fig:figure3}, it can be observed that, all the SI-SDR values of the codecs are negative. This indicates that after passing through the codec, the enhancement performance always deteriorates. For some codecs, such as SpeechTokenizer, which is not trained on noisy data, as the SNR increases, the performance also improves. However, SemantiCodec is limited by the bitrate and shows poor overall performance.

In addition, we also conduct measurements on the real-world dataset in CHiME4. The CHiME4 dataset is divided into four sub-datasets according to the ranges of the approximate SI-SDR values obtained for each audio through TorchAudio-Squim. By analyzing Table \ref{table:table3}, we can conclude that when the TSI-SDR decreases, the codec-reconstructed audio performs worse in terms of signal metrics in the speech enhancement.

\subsubsection{Consistency of SE+ASR Backend}
In this backend, we use ASR to evaluate the loss of enhanced speech integibility. Figure~\ref{fig:figure4} shows that in $\Delta$WER with SNR, at low SNR, the enhancement effect is limited. As SNR increases, the reconstruction-induced enhancement performance loss rises, but it starts to decline once SNR reaches a certain level. By observing each codec's turning points, we find that better reconstruction performance leads to an earlier turning point. Also, when audio is reconstructed before enhancement, the degradation of EnCodec's reconstruction performance increases the enhancement performance loss, as seen from $\Delta$WER with Loudness.

To ensure stability of the backend system, we run it twice using the original audio and measure $\Delta$WER as a baseline. Results show that $\Delta$WER stays low under pure noise, indicating high result certainty of the ASR system under the random sampling strategy.

We further measure the decline of system performance using real-world data. As shown in Table \ref{table:table3}, we do additional tests on the CHiME4 divided by TorchAudio-Squim. 
Although codecs like DAC perform well in reconstruction, the downstream intelligibility performance after reconstruction still has a more than 2$\times$ reduction. 
In addition, when the noise ratio increases, the performance of the system with codec reconstruction decreases significantly. 
This suggests that codec reconstruction does have a significant impact on the downstream speech processing tasks, regardless of its reconstruction quality.

\section{Conclusions}
In this paper, we propose ERSB, a benchmark for evaluating the environmental resilience of neural speech codecs based on reconstruction quality and downstream task consistency. Our experiments show that none of the tested codecs perform well in both aspects. Reconstruction quality varies significantly due to differences in bitrate and techniques, with DAC~\cite{kumar2024high} achieving the best reconstruction performance. However, when evaluated on downstream task consistency, DAC and all other codecs fail to show minimal deviation, indicating that codecs optimized for clean data reconstruction do not necessarily preserve information critical for downstream tasks in complex acoustic environments. These findings highlight a fundamental challenge in codec design and the need for further research to develop environment-resilient codecs.

\section{Acknowledgements}
This work was supported by the China NSFC Project (No. 92370206), the Shanghai Municipal Science and Technology Major Project (2021SHZDZX0102) and the Key Research and Development Program of Jiangsu Province, China (No.BE2022059).

\bibliographystyle{IEEEtran}
\bibliography{mybib}

\begin{thebibliography}{10}
\providecommand{\url}[1]{#1}
\csname url@samestyle\endcsname
\providecommand{\newblock}{\relax}
\providecommand{\bibinfo}[2]{#2}
\providecommand{\BIBentrySTDinterwordspacing}{\spaceskip=0pt\relax}
\providecommand{\BIBentryALTinterwordstretchfactor}{4}
\providecommand{\BIBentryALTinterwordspacing}{\spaceskip=\fontdimen2\font plus
\BIBentryALTinterwordstretchfactor\fontdimen3\font minus \fontdimen4\font\relax}
\providecommand{\BIBforeignlanguage}[2]{{%
\expandafter\ifx\csname l@#1\endcsname\relax
\typeout{** WARNING: IEEEtran.bst: No hyphenation pattern has been}%
\typeout{** loaded for the language `#1'. Using the pattern for}%
\typeout{** the default language instead.}%
\else
\language=\csname l@#1\endcsname
\fi
#2}}
\providecommand{\BIBdecl}{\relax}
\BIBdecl

\bibitem{valle}
S.~Chen, C.~Wang, Y.~Wu \emph{et~al.}, ``{Neural Codec Language Models are Zero-Shot Text to Speech Synthesizers},'' \emph{IEEE Transactions on Audio, Speech and Language Processing}, pp. 1--15, 2025.

\bibitem{wang2024viola}
T.~Wang, L.~Zhou, Z.~Zhang \emph{et~al.}, ``{VioLA}: {Conditional Language Models for Speech Recognition, Synthesis, and Translation},'' \emph{IEEE Transactions on Audio, Speech and Language Processing}, 2024.

\bibitem{yao2025gense}
J.~Yao, H.~Liu, C.~Chen, Y.~Hu, E.~Chng, and L.~Xie, ``Gen{SE}: Generative speech enhancement via language models using hierarchical modeling,'' in \emph{International Conference on Learning Representations}, 2025.

\bibitem{encodec}
A.~D{\'e}fossez, J.~Copet, G.~Synnaeve \emph{et~al.}, ``{High Fidelity Neural Audio Compression},'' \emph{Transactions on Machine Learning Research}, 2023.

\bibitem{kumar2024high}
R.~Kumar, P.~Seetharaman, A.~Luebs \emph{et~al.}, ``{High-Fidelity Audio Compression with Improved RVQGAN},'' \emph{Thirty-Eighth Annual Conference on Neural Information Processing Systems}, vol.~36, 2024.

\bibitem{zhang2024speechtokenizer}
X.~Zhang, D.~Zhang, S.~Li \emph{et~al.}, ``{SpeechTokenizer: Unified Speech Tokenizer for Speech Language Models},'' in \emph{International Conference on Learning Representations}, 2024.

\bibitem{liu2024semanticodec}
H.~Liu, X.~Xu, Y.~Yuan \emph{et~al.}, ``{SemantiCodec}: {An Ultra Low Bitrate Semantic Audio Codec for General Sound},'' \emph{IEEE Journal of Selected Topics in Signal Processing}, pp. 1--14, 2024.

\bibitem{ye2024codec}
Z.~Ye, P.~Sun, J.~Lei \emph{et~al.}, ``{Codec Does Matter: Exploring the Semantic Shortcoming of Codec for Audio Language Model},'' \emph{arXiv preprint arXiv:2408.17175}, 2024.

\bibitem{guo2025review}
Y.~Guo, Z.~Li, H.~Wang, B.~Li, C.~Shao, H.~Zhang, C.~Du, X.~Chen, S.~Liu, and K.~Yu, ``Recent advances in discrete speech tokens: A review,'' \emph{arXiv preprint arXiv:2502.06490}, 2025.

\bibitem{hsu2021hubert}
W.-N. Hsu, B.~Bolte, Y.-H.~H. Tsai \emph{et~al.}, ``{HuBERT: Self-Supervised Speech Representation Learning by Masked Prediction of Hidden Units},'' \emph{IEEE Transactions on Audio, Speech and Language Processing}, vol.~29, pp. 3451--3460, 2021.

\bibitem{du2024cosyvoice}
Z.~Du, Q.~Chen, S.~Zhang \emph{et~al.}, ``{CosyVoice}: {A Scalable Multilingual Zero-Shot Text-to-Speech Synthesizer Based on Supervised Semantic Tokens},'' \emph{arXiv preprint arXiv:2407.05407}, 2024.

\bibitem{wu2024codec}
H.~Wu, H.-L. Chung, Y.-C. Lin, Y.-K. Wu, X.~Chen, Y.-C. Pai, H.-H. Wang, K.-W. Chang, A.~H. Liu, and H.-y. Lee, ``Codec-{SUPERB}: An in-depth analysis of sound codec models,'' \emph{arXiv preprint arXiv:2402.13071}, 2024.

\bibitem{mousavi2024dasb}
P.~Mousavi, L.~Della~Libera, J.~Duret \emph{et~al.}, ``{DASB}--{Discrete Audio and Speech Benchmark},'' \emph{arXiv preprint arXiv:2406.14294}, 2024.

\bibitem{dns5}
H.~Dubey, A.~Aazami, V.~Gopal, B.~Naderi, S.~Braun, R.~Cutler, H.~Gamper, M.~Golestaneh, and R.~Aichner, ``{ICASSP} 2023 deep noise suppression challenge,'' in \emph{International Conference on Acoustics, Speech, and Signal Processing}.\hskip 1em plus 0.5em minus 0.4em\relax IEEE, 2023.

\bibitem{TorchAudio-Squim}
A.~Kumar, K.~Tan, Z.~Ni, P.~Manocha, X.~Zhang, E.~Henderson, and B.~Xu, ``{TorchAudio-Squim}: Reference-less speech quality and intelligibility measures in torchaudio,'' in \emph{International Conference on Acoustics, Speech, and Signal Processing}.\hskip 1em plus 0.5em minus 0.4em\relax IEEE, 2023.

\bibitem{chime1}
J.~Barker, E.~Vincent, N.~Ma, C.~Christensen, and P.~Green, ``The {PASCAL} {CHiME} speech separation and recognition challenge,'' \emph{Computer Speech and Language}, vol.~27, no.~3, pp. 621--633, 2013.

\bibitem{dns1}
C.~K. Reddy, V.~Gopal, R.~Cutler, E.~Beyrami, R.~Cheng, H.~Dubey, S.~Matusevych, R.~Aichner, A.~Aazami, S.~Braun \emph{et~al.}, ``The {INTERSPEECH} 2020 deep noise suppression challenge: Datasets, subjective testing framework, and challenge results,'' in \emph{Proc. ISCA Interspeech}, 2020.

\bibitem{vincent20164th}
E.~Vincent, S.~Watanabe, J.~Barker, and R.~Marxer, ``The 4th {CHiME} speech separation and recognition challenge,'' \emph{URL: http://spandh. dcs. shef. ac. uk/chime\_challenge/(last accessed on 1 August, 2018)}, 2016.

\bibitem{whisper}
A.~Radford, J.~W. Kim, T.~Xu \emph{et~al.}, ``{Robust Speech Recognition via Large-Scale Weak Supervision},'' in \emph{International Conference on Machine Learning}.\hskip 1em plus 0.5em minus 0.4em\relax PMLR, 2023, pp. 28\,492--28\,518.

\bibitem{subakan2021attention}
C.~Subakan, M.~Ravanelli, S.~Cornell, M.~Bronzi, and J.~Zhong, ``Attention is all you need in speech separation,'' in \emph{International Conference on Acoustics, Speech, and Signal Processing}.\hskip 1em plus 0.5em minus 0.4em\relax IEEE, 2021, pp. 21--25.

\bibitem{wham}
G.~Wichern, J.~Antognini, M.~Flynn, L.~R. Zhu, E.~McQuinn, D.~Crow, E.~Manilow, and J.~L. Roux, ``{WHAM}!: Extending speech separation to noisy environments,'' in \emph{Proc. ISCA Interspeech}, 2019, pp. 1368--1372.

\bibitem{shi2024versa}
J.~Shi, H.-j. Shim, J.~Tian \emph{et~al.}, ``{VERSA: A Versatile Evaluation Toolkit for Speech, Audio, and Music},'' \emph{arXiv preprint arXiv:2412.17667}, 2024.

\end{thebibliography}

\end{document}